\documentclass[twocolumn,secnumarabic,footinbib,aps,pra,superscriptaddress]{revtex4-1}
\usepackage{biolinum}
\usepackage[latin9]{inputenc}
\usepackage{geometry}
\usepackage{color}
\usepackage{calc}
\usepackage{bbold}
\usepackage{float}
\usepackage{textcomp}
\usepackage{amsmath}
\usepackage{amssymb}
\usepackage{wasysym}
\usepackage{longtable}
\usepackage{stackrel}
\usepackage{lipsum}
\usepackage{multirow}
\usepackage{graphicx}
\usepackage{xcolor}
\PassOptionsToPackage{normalem}{ulem}
\usepackage{ulem}

\begin{document}

\title{An Experimentally Verified Approach to non-Entanglement-Breaking Channel Certification}

\author{Yingqiu Mao}
\thanks{These authors contributed equally to this work.}
\affiliation{Hefei National Laboratory for Physical Sciences at Microscale and Department of Modern Physics, University of Science and Technology of China, Hefei, Anhui 230026, P.~R.~China}
\affiliation{CAS Center for Excellence and Synergetic Innovation Center in Quantum Information and Quantum Physics, University of Science and Technology of China, Hefei, Anhui 230026, P.~R.~China}

\author{Yi-Zheng Zhen}
\thanks{These authors contributed equally to this work.}
\affiliation{Institute for Quantum Science and Engineering, Southern University of Science and Technology, Shenzhen, Guangdong 518055, P.~R.~China}
\affiliation{Hefei National Laboratory for Physical Sciences at Microscale and Department of Modern Physics, University of Science and Technology of China, Hefei, Anhui 230026, P.~R.~China}

\author{Hui Liu}
\affiliation{Hefei National Laboratory for Physical Sciences at Microscale and Department of Modern Physics, University of Science and Technology of China, Hefei, Anhui 230026, P.~R.~China}
\affiliation{CAS Center for Excellence and Synergetic Innovation Center in Quantum Information and Quantum Physics, University of Science and Technology of China, Hefei, Anhui 230026, P.~R.~China}

\author{Mi Zou}
\affiliation{Hefei National Laboratory for Physical Sciences at Microscale and Department of Modern Physics, University of Science and Technology of China, Hefei, Anhui 230026, P.~R.~China}
\affiliation{CAS Center for Excellence and Synergetic Innovation Center in Quantum Information and Quantum Physics, University of Science and Technology of China, Hefei, Anhui 230026, P.~R.~China}

\author{Qi-Jie Tang}
\affiliation{Hefei National Laboratory for Physical Sciences at Microscale and Department of Modern Physics, University of Science and Technology of China, Hefei, Anhui 230026, P.~R.~China}
\affiliation{CAS Center for Excellence and Synergetic Innovation Center in Quantum Information and Quantum Physics, University of Science and Technology of China, Hefei, Anhui 230026, P.~R.~China}

\author{Si-Jie Zhang}
\affiliation{Hefei National Laboratory for Physical Sciences at Microscale and Department of Modern Physics, University of Science and Technology of China, Hefei, Anhui 230026, P.~R.~China}
\affiliation{CAS Center for Excellence and Synergetic Innovation Center in Quantum Information and Quantum Physics, University of Science and Technology of China, Hefei, Anhui 230026, P.~R.~China}

\author{Jian Wang}
\affiliation{Hefei National Laboratory for Physical Sciences at Microscale and Department of Modern Physics, University of Science and Technology of China, Hefei, Anhui 230026, P.~R.~China}
\affiliation{CAS Center for Excellence and Synergetic Innovation Center in Quantum Information and Quantum Physics, University of Science and Technology of China, Hefei, Anhui 230026, P.~R.~China}

\author{Hao Liang}
\affiliation{Hefei National Laboratory for Physical Sciences at Microscale and Department of Modern Physics, University of Science and Technology of China, Hefei, Anhui 230026, P.~R.~China}
\affiliation{CAS Center for Excellence and Synergetic Innovation Center in Quantum Information and Quantum Physics, University of Science and Technology of China, Hefei, Anhui 230026, P.~R.~China}

\author{Weijun Zhang}
\affiliation{State Key Laboratory of Functional Materials for Informatics, Shanghai Institute of Microsystem and Information Technology, Chinese Academy of Sciences, Shanghai 200050, P.~R.~China}

\author{Hao Li}
\affiliation{State Key Laboratory of Functional Materials for Informatics, Shanghai Institute of Microsystem and Information Technology, Chinese Academy of Sciences, Shanghai 200050, P.~R.~China}

\author{Lixing You}
\affiliation{State Key Laboratory of Functional Materials for Informatics, Shanghai Institute of Microsystem and Information Technology, Chinese Academy of Sciences, Shanghai 200050, P.~R.~China}

\author{Zhen Wang}
\affiliation{State Key Laboratory of Functional Materials for Informatics, Shanghai Institute of Microsystem and Information Technology, Chinese Academy of Sciences, Shanghai 200050, P.~R.~China}

\author{Li Li}
\affiliation{Hefei National Laboratory for Physical Sciences at Microscale and Department of Modern Physics, University of Science and Technology of China, Hefei, Anhui 230026, P.~R.~China}
\affiliation{CAS Center for Excellence and Synergetic Innovation Center in Quantum Information and Quantum Physics, University of Science and Technology of China, Hefei, Anhui 230026, P.~R.~China}

\author{Nai-Le Liu}
\affiliation{Hefei National Laboratory for Physical Sciences at Microscale and Department of Modern Physics, University of Science and Technology of China, Hefei, Anhui 230026, P.~R.~China}
\affiliation{CAS Center for Excellence and Synergetic Innovation Center in Quantum Information and Quantum Physics, University of Science and Technology of China, Hefei, Anhui 230026, P.~R.~China}

\author{Kai Chen}
\email{	kaichen@ustc.edu.cn}
\affiliation{Hefei National Laboratory for Physical Sciences at Microscale and Department of Modern Physics, University of Science and Technology of China, Hefei, Anhui 230026, P.~R.~China}
\affiliation{CAS Center for Excellence and Synergetic Innovation Center in Quantum Information and Quantum Physics, University of Science and Technology of China, Hefei, Anhui 230026, P.~R.~China}

\author{Teng-Yun Chen}
\email{tychen@ustc.edu.cn}
\affiliation{Hefei National Laboratory for Physical Sciences at Microscale and Department of Modern Physics, University of Science and Technology of China, Hefei, Anhui 230026, P.~R.~China}
\affiliation{CAS Center for Excellence and Synergetic Innovation Center in Quantum Information and Quantum Physics, University of Science and Technology of China, Hefei, Anhui 230026, P.~R.~China}

\author{Jian-Wei Pan}
\email{pan@ustc.edu.cn}
\affiliation{Hefei National Laboratory for Physical Sciences at Microscale and Department of Modern Physics, University of Science and Technology of China, Hefei, Anhui 230026, P.~R.~China}
\affiliation{CAS Center for Excellence and Synergetic Innovation Center in Quantum Information and Quantum Physics, University of Science and Technology of China, Hefei, Anhui 230026, P.~R.~China}

\begin{abstract}
Ensuring the non-entanglement-breaking (non-EB) property of quantum channels is crucial for the effective distribution and storage of quantum states. 
However, a practical method for direct and accurate certification of the non-EB feature is highly desirable. 
Here, we propose and verify a realistic source based measurement device independent certification of non-EB channels. 
Our method is resilient to repercussions on the certification from experimental conditions, such as multiphotons and imperfect state preparation, and can be implemented with information incomplete set. 
We achieve good agreement between experimental outcomes and theoretical predictions, which is validated by the expected results of the ideal semi-quantum signaling game, and accurately certify the non-EB channels. 
Furthermore, our approach is highly robust to effects from noise. 
Therefore, the proposed approach can be expected to play a significant role in the design and evaluation of realistic quantum channels. 

\end{abstract}

\maketitle

Numerous quantum information tasks have shown better performance than their classical counterparts, when the entanglement \cite{horodecki2009RMP,GUHNE20091,Friis2019} between quantum states for the corresponding quantum process is maintained \cite{nielsen2002quantum,Giovannetti2004,Scarani2009RMP}. 
Notably, effective entanglement distribution is a crucial precondition for unconditional security in quantum cryptography \cite{Scarani2009RMP,curty2004}, while persisting entanglement over computation time is necessary for the speed-up of quantum computing \cite{jozsa2003role}. 
Such processes require at least the participating channel to be non-entanglement-breaking (non-EB), i.e., the channel guarantees non-vanishing entanglement when a party of an entangled pair transmits through it \cite{horodecki2003entanglement}. 
In light of the growing importance of quantum networks, and the various ways in which real-life quantum channels are implemented, it is desirable to search for a practical, general approach to certify non-EB channels, and guide the design and evaluation of quantum channels.

Obviously, one may in principle certify non-EB channels with full device-independence, if one sends one party of an entangled pair through the channel and measures the output bipartite states using a loophole-free Bell test \cite{brunner2014bell}. 
However, this method certifies non-locality, which is a different resource from entanglement \cite{Brunner2005NJP} and requires much stricter experimental conditions than entanglement verification \cite{hensen2015loophole,giustina2015significant,shalm2015strong}. 
Even though one can replace the Bell test with various kinds of entanglement witnesses \cite{horodecki2009RMP,GUHNE20091,Buscemi2012cmp,buscemi2012sqg,Branciard2013MDIEW,mdiew,Nawareg2015,Verbanis2016MDIEW,Supic2017,Bowles2018}, it is still difficult to lower the experimental requirements, due to the need for a near-perfect maximally entangled state source. 
Thus, this method is rarely seen in practical applications, where one usually sends single-photon states directly through the channel, and performs quantum process tomography to determine the exact process of a quantum channel \cite{1997PRLtomo,2001PRLtomo,altepeter2003ancilla,o2004quantum}. 
Still, imperfect detection devices may cause reconstruction of nonphysical states \cite{Schwemmer2015}, which leads to wrong characterizations of the channel, and in some adversarial situations, may even lead to security loopholes \cite{zhao2008pra,lydersen2010hacking,gerhardt2011NC,Weier2011NJP}. 
Therefore, it is vital for the design and implementation of realistic quantum channels to find a practical and efficient approach for non-EB certification.

\begin{figure}[!hbt]\centering
\includegraphics[scale=0.65,trim={0.6cm 0.5cm 0.5cm 0.5cm},clip]{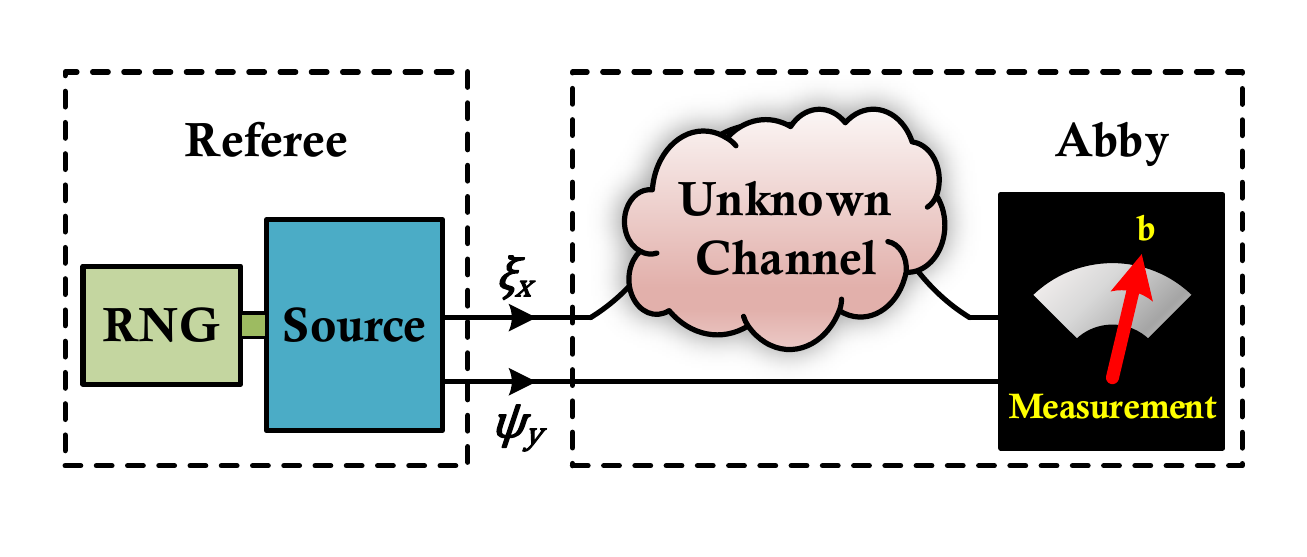}
\caption{\label{fig0:game}
Schematics of the ideal SQSG \cite{rosset2018resource}. 
Referee first asks a random quantum question $\xi_x$ to the player Abby, who inputs $\xi_x$ to an unknown channel $\mathcal{N}$ to be certified. 
Later, referee randomly asks another quantum question $\psi_y$.
Then, Abby feeds the channel output and $\psi_y$ into an untrusted measurement device, which yields an answer $b$. 
Based on  $\xi_x$, $\psi_y$, and $b$, referee calculates an average payoff, and Abby wins the game if it is larger than $0$.
RNG: random number generator.}
\end{figure}

Recently, Rosset \textit{et al.} \cite{rosset2018resource} proposed a theoretical solution to these problems for certifying non-EB channels. 
By playing a simple semi-quantum signaling game (SQSG), the non-EB channel may be proven as a necessary resource to win (see Fig.~\ref{fig0:game}) through a violation of an inequality. 
This SQSG method theoretically verifies the non-EB channel in the measurement-device-independent (MDI) scenario, which can be robust to detection errors and generalized to other scenarios \cite{Guerini2019}.
Unfortunately, SQSG is based on single-copy, ideally prepared quantum states that belong to an information complete set \cite{Wilde2013}, which has led to its correctness not yet verified by experiments. 
To experimentally test the SQSG, multiphoton emissions are unavoidable in realistic sources.
Such sources will not only bring about security loopholes \cite{PNS2000}, but may also reduce the certification efficiency. 
Therefore, it is necessary to develop a reliable and experimentally verifiable approach to certify quantum channels based on realistic quantum states. 
If the theory can be generalized to practical sources and rigorously verified with a credible experiment, it is also crucial to consider important problems such as inaccurate certification when the quantum state preparation is imperfect, and whether less states can be used instead of the information complete set.

In this Letter, we propose and experimentally demonstrate a general and practical approach for certifying non-EB channels. 
Based on the ideal SQSG, we develop an experimentally verifiable approach which does not rely on perfectly prepared, single-photon states. 
Then, we precisely design and realize a stable weak coherent pulse (WCP) based fiber-type experimental system with non-EB strength controllable typical quantum channels. 
We demonstrate the non-EB certification, and the results indicate good accordance between the experimental statistics and our theory, which are further confirmed by the predictions of the ideal SQSG. 
Moreover, our approach does not require perfect information-complete state preparation, and is highly robust to noise.

\textit{Realistic source based MDI non-EB certification} -- 
To describe how much Abby will win the SQSG, an average payoff has been given by Rosset \textit{et al.} \cite{rosset2018resource}:
\begin{equation}
I_\mathcal{N}=\sum_{x,y,b}\wp(b,x,y)P_\mathcal{N}\left(b|\xi_x,\psi_y\right),
\label{eq:payoff}
\end{equation}
where $\wp(b,x,y)$ is the payoff function and $P_\mathcal{N}\left(b|\xi_x,\psi_y\right)$ is the probability of Abby outputting answer $b$ by jointly measuring $\mathcal{N}\left(\xi_x\right)$ and $\psi_y$ (see Fig.~\ref{fig0:game}).
In the ideal SQSG, referee is required to use only single-copy, perfectly prepared states of $\xi_x$ and $\psi_y$, which are restricted to an information complete set \cite{rosset2018resource}.
When Abby performs the joint measurement, referee can obtain $I_{\mathcal{N}_{EB}}\leqslant0$ for any EB channel. 
As a result, one can certify the non-EB channel with a positive $I_\mathcal{N}$. 
In this work, we focus on Eq.~(\ref{eq:payoff}) with $\wp\left(b\neq0,x,y\right)=0$.

If one weakens the assumption on the referee, such that he only has full knowledge of the states $\xi_x$ and $\psi_y$, which may not necessarily form an information complete set, then, $I_\mathcal{N}$ for EB channels can be bounded as
\begin{equation}
\begin {aligned}
\label{eq:prac-bound}
C_{\textrm{EB}}&=\max_{\mathcal{N}_{\textrm{EB}}} \sum_{x,y}\wp(0,x,y)P_{\mathcal{N}_{\textrm{EB}}}\left(0|\xi_x,\psi_y\right)\\
&=d^2 \max_{\omega_{\textrm{sep}}}{\rm tr}\left[W\omega_{\textrm{sep}}\right],
\end{aligned}
\end{equation}
where $W=\sum_{xy}\wp\left(0,x,y\right)\xi_x^T\otimes\psi_y^T$, and $\omega_{\textrm{sep}}$ is a separable state. By adopting the experimental bound $C_{\textrm{EB}}$, one can use the inequality $I_{\cal N}>C_{\textrm{EB}}$ as a certification for non-EB channels under realistic conditions.

To exclude effects from multiphoton emissions of realistic sources \cite{PNS2000}, we use the decoy-state technique to obtain $I_\mathcal{N}$ contributed by single-photon events only \cite{hwang2003quantum,wang2005beating,lo2005decoy}. 
We consider phase-randomized WCPs, which is one of the most common sources in experiments. 
The photons follow the Poisson distribution, i.e., $\rho_\alpha=e^{-\alpha}\sum_{n=0}^{\infty}\alpha^n\left|n\right\rangle\left\langle n\right|/n!$, where $\alpha$ is the mean photon number per pulse and $n$ is the photon number. When pulses $\xi_x$ and $\psi_y$ are prepared with intensities $\alpha_\xi$ and $\alpha_\psi$, respectively, the probability of Abby obtaining the answer $b$ by joint measurement may be defined as the following gain \cite{Xu2013practical}
\begin{equation}
\label{eq:gain}
Q_{b,\xi_x,\psi_y}^{\alpha_\xi \alpha_\psi} = e^{-\alpha_\xi-\alpha_\psi}\sum_{n,m=0}^{\infty}\frac{\alpha_\xi^n\alpha_\psi^m}{n!m!}Y_{b,\xi_x,\psi_y}^{nm},
\end{equation}
where $Y_{b,\xi_x,\psi_y}^{nm}$ is the conditional probability of detection event $b$, given that  $n$-photon and  $m$-photon pulses are emitted in $\xi_x$ and $\psi_y$, respectively. 
When mean photon numbers of $\xi_x$ and $\psi_y$ pulses are randomly selected among three different values, i.e., decoy states $\alpha_\xi,\alpha_\psi\in\left\{\mu,\nu,\omega\right\}$ with $\mu>\nu>\omega$, $P_\mathcal{N}\left(b|\xi_x,\psi_y\right)$, or equivalently $Y_{b,\xi_x,\psi_y}^{11}$, can be determined. 
From linear equations of different gains $Q_{b,\xi_x,\psi_y}^{\alpha_\xi \alpha_\psi}$, $Y_{b,\xi_x,\psi_y}^{11}$ can be lower and upper bounded, denoted by $Y_{b,\xi_x,\psi_y}^{11,L}$ and $Y_{b,\xi_x,\psi_y}^{11,U}$, respectively. Consequently, $I_{\cal N}$ in our work has a lower bound
\begin{equation}
\label{eq:lb-payoff}
I_{\cal N}\geqslant I_{\cal N}^L = \sum_{x,y}\wp\left(0,x,y\right)Y_{0,\xi_x,\psi_y}^{11,L/U},
\end{equation}
where $Y_{0,\xi_x,\psi_y}^{11,L}$ or $Y_{0,\xi_x,\psi_y}^{11,U}$ are chosen according to the sign of $\wp\left(0,x,y\right)$.

Then, under realistic conditions, one can apply the inequality
\begin{equation}
\label{eq:fin-ineq}
I_{\cal N}^L>C_{\textrm{EB}},
\end{equation}
to analyze the non-EB features of a tested channel. Thus, we obtain an experimentally verifiable, realistic source based MDI (RS-MDI) non-EB channel certification. Moreover, Eq.~(\ref{eq:fin-ineq}) is only related to detection events caused by single-photon emissions of the source, making our approach robust to multiphoton components. We leave the theoretical details 
 to the Supplemental Material \footnote{See Supplemental Material for details}.

\begin{figure*}[t]\centering
\includegraphics[scale=0.43]{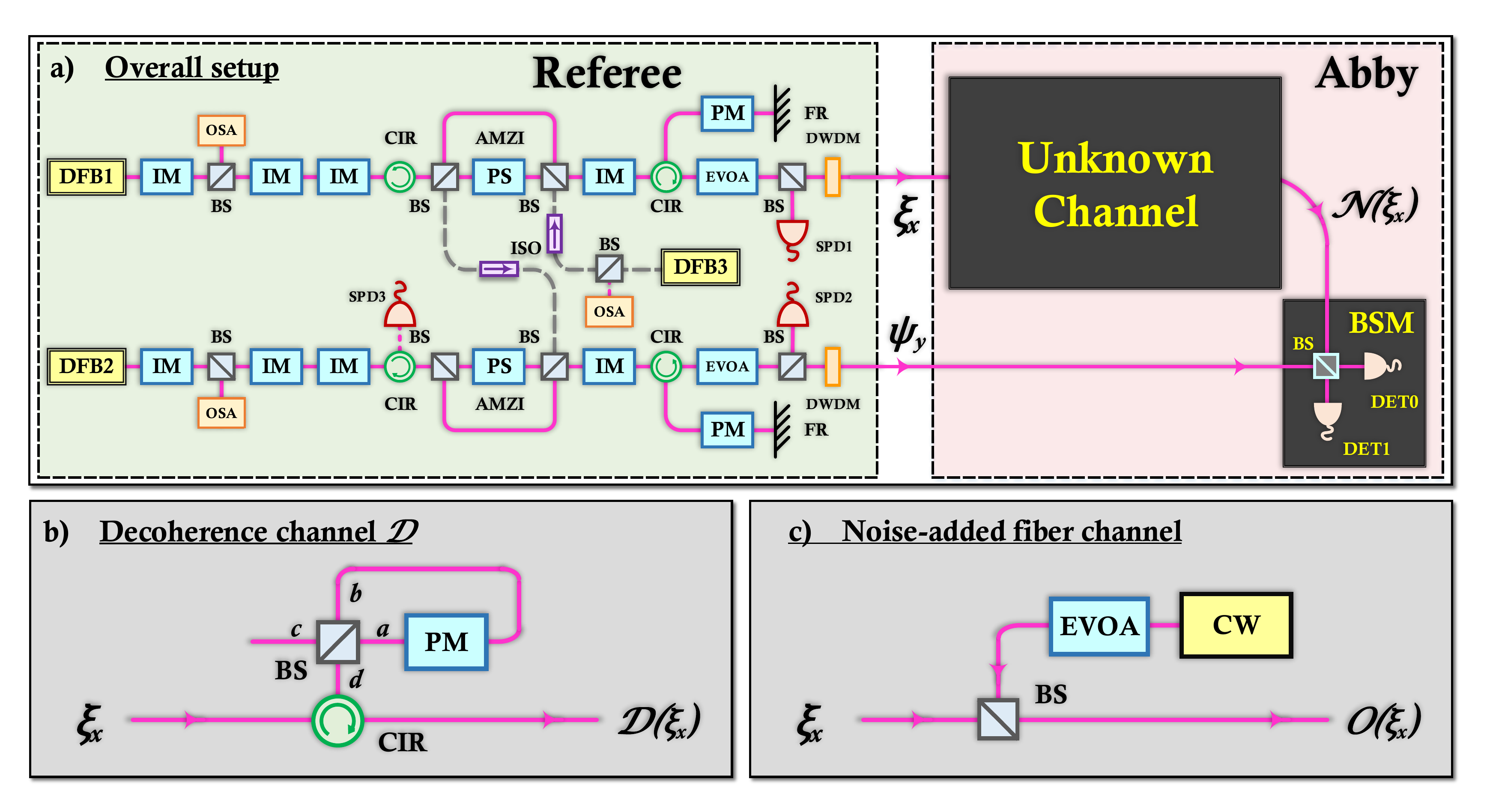}
\caption{\label{fig1:setup}
Setup of RS-MDI certification for non-EB channels. 
DFB: distributed feedback laser; IM: intensity modulator; AMZI: asymmetric Mach-Zehnder interferometer; BS: beam splitter; CIR: circulator; PS: phase shifter; PM: phase modulator; FR: fiber reflector; EVOA: electronic variable optical attenuator; DWDM: dense wavelength-division multiplexer; SPD: InGaAs gated single photon detector; CW: continuous wave laser; DET0/DET1: superconducting nanowire single photon detectors (SNSPD). 
}
\end{figure*}

\textit{Experiment} --- 
To verify the feasibility of our
 method, it is necessary to design and realize an EB strength controllable, stable experimental system. 
Without loss of generality, 
we design a full polarization maintaining fiber verification system (see Fig.~\ref{fig1:setup}).
The referee has two identical state-preparation modules, and by using time-bin and phase encoding \cite{ma2012alternative}, he sends WCPs of $\xi_x$ and $\psi_y$ to Abby. 
$\xi_x$ and $\psi_y$ are randomly selected from the eigenstates of the three Pauli matrices,
i.e., encoded as the first and second time-bins for $Z$ basis, and encoded in the relative phase between the two time-bins for $X$ ($Y$) basis.

Our experimental setup is composed of three portions: state preparation, detection, and the channel to be tested. 
For state preparation,
time-bin states are created using an AMZI 
, and the basis of $Z$ or $X$($Y$) is chosen with the following IM. 
Phase states are created using the FR, PM, and CIR.
The pulses are lowered down to single-photon level with an EVOA, and are filtered with a $100$ GHz narrow pass-band filter for spectral noise. 
Based on the tomography of $\xi_x$ and $\psi_y$ \cite{Sun2016}, the experimental bound $C_{\textrm{EB}}$ can be calculated with Eq.~(\ref{eq:prac-bound}).

The state detection is implemented with a partial Bell-state measurement (BSM). 
When coincidence counts occur at two alternative time bins of Det0 and Det1, projection on $\left|\Psi^{-}\right\rangle=\left(\left|01\right\rangle -\left|10\right\rangle \right)/\sqrt{2}$ is selected, which is labeled $b=0$, and the gain in Eq.~(\ref{eq:gain}) can be determined. 

In general quantum information tasks, the decoherence of quantum states is one of the main causes for the channel to destroy entanglement.
Therefore, we construct a fiber-type Sagnac interformeter based channel to be tested (see Fig.~\ref{fig1:setup}b), where the strength of channel decoherence, $\gamma$, is precisely controlled through varying the voltage of the PM in the interformeter.
The coherence is suppressed when $\gamma$ increases, and the channel becomes completely EB iff $\gamma=1$.
Additionally, as noise is one of the most important factors affecting the performance in non-EB channel based practical applications 
 \cite{townsend1997simultaneous,frohlich2017long,Mao2018QKD}, 
for simplicity and without loss of generality, we implement a test fiber channel (see Fig.~\ref{fig1:setup}c) to study the effects of noise on our approach. 
We leave the experimental details
in the Supplemental Material \cite{Note1}.

\begin{figure}\centering
\includegraphics[width=\linewidth]{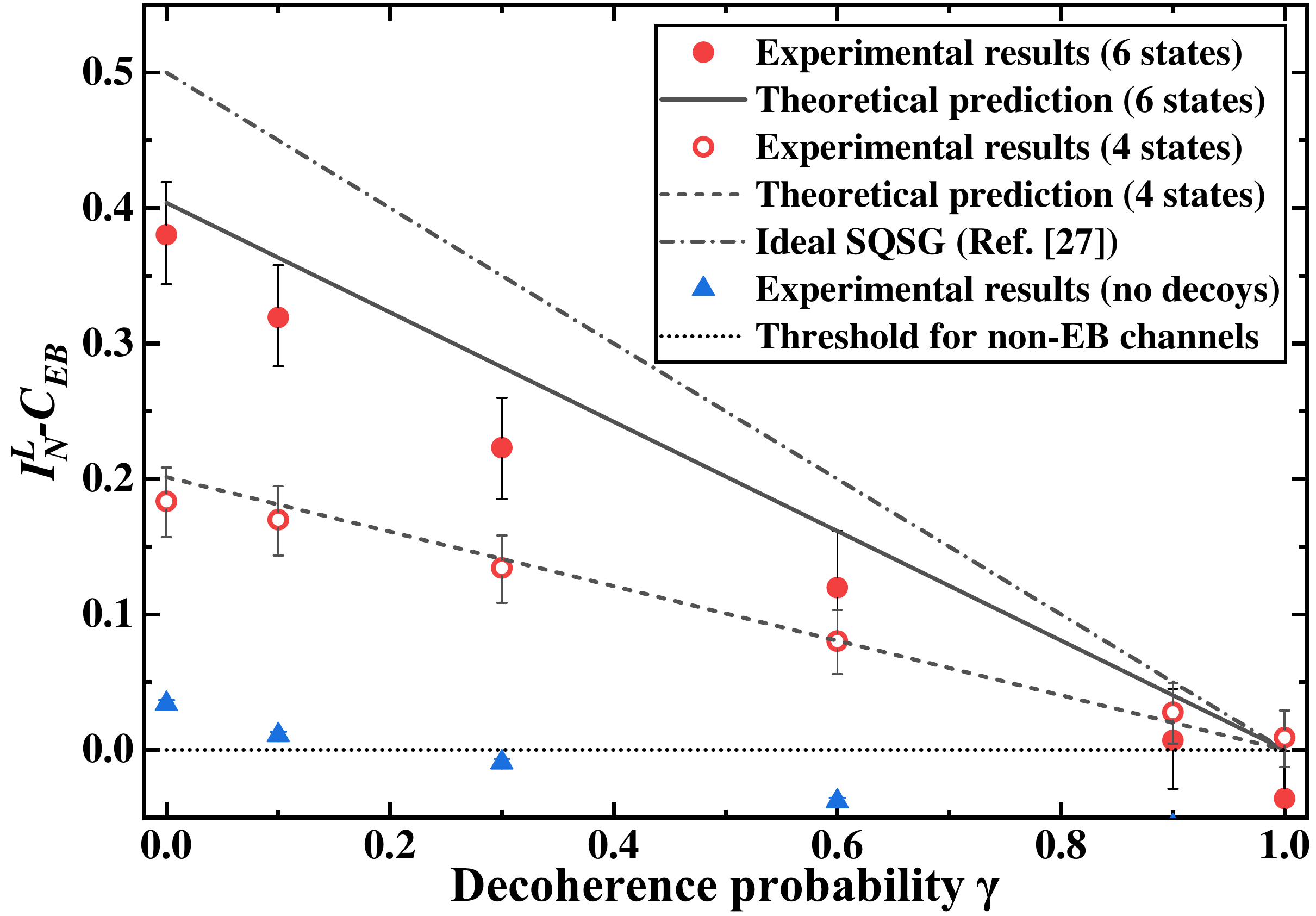}
\caption{\label{fig:decoherence}
Certification results for the decoherence channel. 
The black bars refer to statistical fluctuations of 1 standard deviation. The experimental results and theoretical prediction are obtained with our method.
}
\end{figure}

\textit{Results and Discussion} --- 
By varying the decoherence strength $\gamma$ of the channel to be tested (Fig. \ref{fig1:setup}b), we first verify the correctness of our method. 
Using the six states of $\xi_x$ and $\psi_y$ as an information complete set, we 
obtain $I_{\cal N}^L$ using Eq.~(\ref{eq:lb-payoff}) for each $\gamma$. Results are shown as the red dots in Fig. \ref{fig:decoherence}, which indicates that the non-EB regions can be accurately certified.
For $\gamma=1$, $I_{\cal N}^L$ is $0.011$, which does not violate the experimental bound $C_{\textrm{EB}}=0.047$, and is 
in accordance to the fact that the fully decoherence channel is EB. 
Particularly, if the ideal SQSG bound $0$ is directly applied
 , an incorrect certification will occur. 
 Thus, experimental results show the necessity to correct the EB bound considering imperfect state preparation, and the practical value of our approach.

The experimental results are completely consistent with our RS-MDI non-EB certification theory (black solid line, Fig.~\ref{fig:decoherence}).
From Eq.~(\ref{eq:lb-payoff}), we see that because $I_{\cal N}^L$ lower bounds the theoretical predictions, the measured results are all below the black solid line.
In principle, if infinite sets of decoys are used, it can be expected that the two will coincide \cite{ma2005practical}. 
The predictions of decoherence channel with ideal SQSG \cite{rosset2018resource} is also shown (black dash-dot line, Fig.~\ref{fig:decoherence}). 
It can be seen that our theoretical and experimental results are both consistent with the 
 predictions of the ideal SQSG, but the results of the former are slightly lower than the latter. 
This is due to the fact that imperfect state preparation is allowed in our approach. 
This small decrease in $I_{\cal N}$ value is acceptable, as our RS-MDI approach confirms the non-EB feature of tested channels under practical conditions. 
Through comparison with predictions of the ideal SQSG, the correctness of our approach is validated.

In addition, we show the necessity of applying the decoy-state technique for practical sources.
Without such a technique, i.e., directly applying the gain 
in Eq.~(\ref{eq:gain}) into Eq.~(\ref{eq:payoff}), 
the performance of the certification is severely damaged (
blue triangles, Fig.~\ref{fig:decoherence}).
Here, only channels of $\gamma\in\left\{0,0.1\right\}$ can be certified. 
This is due to the fact that most of WCPs are vacuum and multi-photon emissions, 
successful BSM 
 events $b=0$ 
 are sharply reduced.
Also, multiphoton emissions cause high errors in detection events for $X$ and $Y$ basis \cite{wang2018enabling}, resulting in significant decrease of the overall average payoff.
It is the application of the decoy-state technique that removes detection events from vacuum and multiphoton emissions, and strictly bounds the probability of single-photon detection events, so that the values $I_{\cal N}^L$ can be accurately determined, ensuring correct certification of the non-EB feature for the tested channel.

Furthermore, to reduce experimental resources and complexity,
we demonstrate our approach using fewer states. 
By reducing 
 to four states (eigenstates of $Z$ and $Y$) of $\xi_x$ and $\psi_y$, the above experiment is repeated, with results shown as the red circles in Fig.~\ref{fig:decoherence}. 
Although the values of $I_{\cal N}^L$ have slightly decreased, 
it can be seen that the experimental results follow our theoretical predictions well, and that the behavior of $I_{\cal N}^L$  to $\gamma$ is the same as that with six states.
Non-EB channels from $0\leqslant\gamma\leqslant0.9$ can still be certified. 
Thus, it can be seen that our method relaxes the requirement of information complete set, and can certify non-EB channels with less resources. 

\begin{figure}\centering
\includegraphics[width=\linewidth]{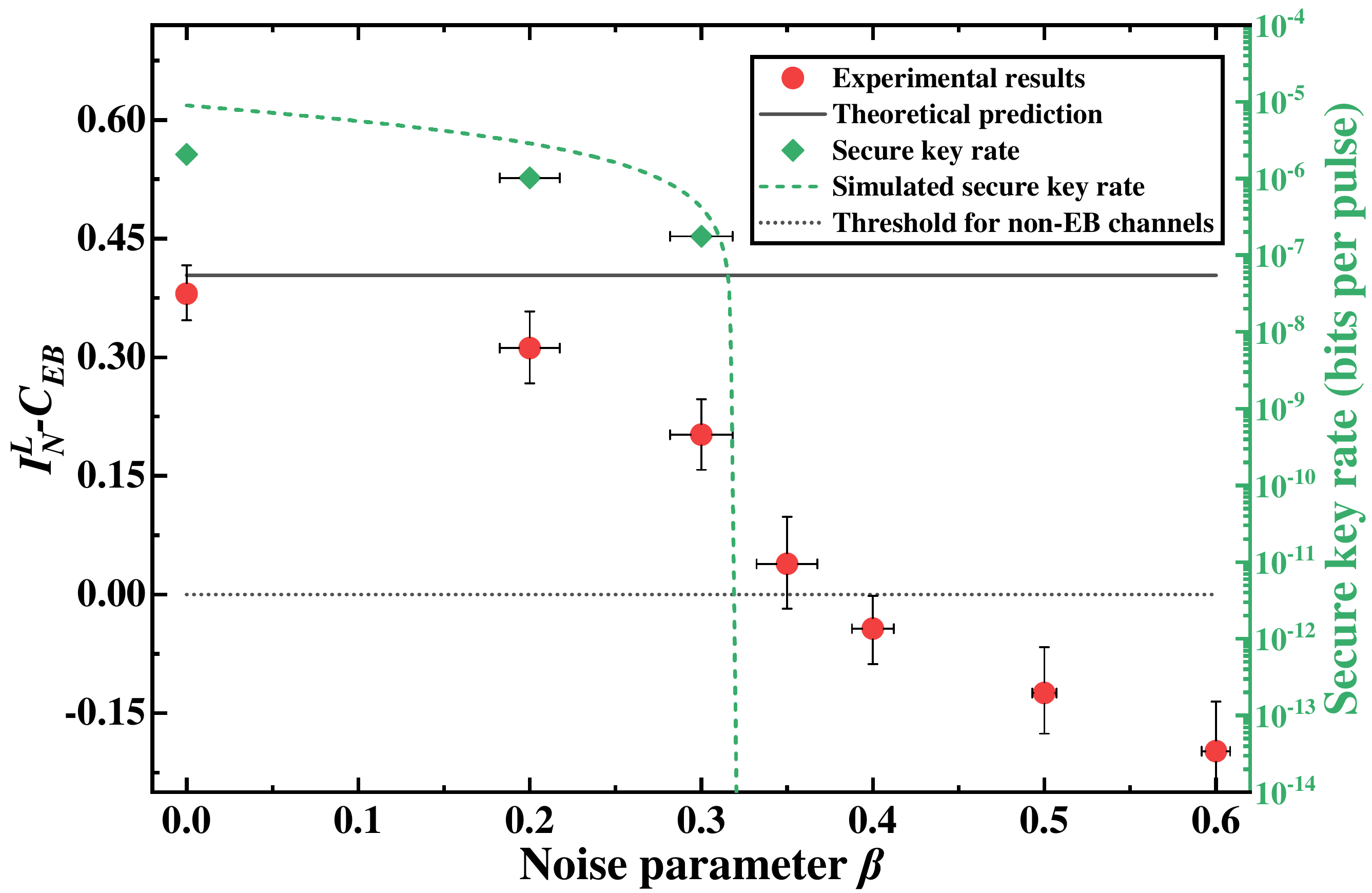}
\caption{\label{fig3:noise}
Certification results for the noise-added fiber channel. 
The black bars refer to statistical fluctuations of 1 standard deviation. 
The experimental results and theoretical prediction are obtained with our method. The black solid line represents the certification when no noise is added for the fiber channel.
}
\end{figure}

Finally, using the channel shown in Fig.~\ref{fig1:setup}(c), and altering the strength of noise $\beta$, we investigate the effects of noise on our method.
For each value of $\beta$, the corresponding average payoff is obtained,  shown as the red dots in Fig.~\ref{fig3:noise}. 
It can be seen that the results monotonically decrease with the increase of $\beta$.
For $\beta\leqslant0.35$, 
our method can certify the noise-added channel non-EB. 
For $\beta>0.35$, $I_{\cal N}^L<C_{EB}$ and the non-EB feature of the tested channel is not confirmed. 
For a simple fiber channel 
(i.e., the identity channel), the noise limit is $~35\%$ of the signal photons.

Due to the fact that the non-EB channel is a necessary precondition for quantum key distribution (QKD) \cite{curty2004}, this requirement can be used to verify the correctness of our method under the influence of noise.
With the same $Q_{b,\xi_x,\psi_y}^{\alpha_\xi \alpha_\psi}$ and $Y_{b,\xi_x,\psi_y}^{11}$ in the RS-MDI non-EB channel certification, we calculate the key rates for standard 4-state MDI-QKD 
\cite{mdiqkd}, with experimental key rates (green diamonds) and simulation (dashed line) shown in Fig.~\ref{fig3:noise}.
Secure keys are 
 generated for 
  $\beta\in\left\{0, 0.2, 0.3\right\}$, which confirm the non-EB feature of the channel certified by our method. 
Although no keys are generated for $\beta=0.35$, this 
 may be fixed by extending standard MDI-QKD to 6 states and further optimizing the intensity and number of decoy states. 
Therefore, our method is verified to tolerate a certain degree of noise, indicating strong practicability.

\textit{Conclusions} --- 
To overcome the difficulties for accurate and practical certification of the non-EB property of quantum channels, we have proposed and verified a RS-MDI approach, based on the ideal SQSG and considering realistic experimental conditions. 
Our method does not require perfectly prepared quantum states from a certain set, can avoid effects from multiphotons, and enjoys the advantages of MDI. 
We have designed a stable and precise experimental system with EB strength controllable typical channels, and successfully implemented our method for non-EB channel certification. 
By using only decoy-state assisted WCPs, an arbitrary set of quantum states, and an experimental bound, accurate certification of non-EB channels is achieved, which are also validated by the expected results of the ideal SQSG. 
Furthermore, robustness against noise of our approach is observed and justified. 
Therefore, our approach can be expected to play a significant role in benchmarking functions of realistic quantum devices such as quantum memories and quantum gates, and is a step forward in bridging the gap between theory and practice for justifying quantum advantages of novel quantum technologies.

\subsection{Acknowledgments}

Y. M. and Y.-Z. Z. especially thank Prof. Francesco Buscemi for numerous advice and encouragement throughout the project.
We thank Jun Zhang, Wen-Yuan Wang, Yan-Lin Tang, Ping Xu, Leonardo Guerini, and Qinghe Mao for valuable and illuminating discussions.

After submission, we became aware that a similar experiment was performed using a different type of system in \cite{Graffitti2019}. The scenario considered in our work can be further generalized to the semi-quantum prepare-and-measure scenario \cite{Guerini2019}.

This work has been supported by 
the National Key R\&D Program of China (2017YFA0303903),
the Chinese Academy of Science, 
the National Fundamental Research Program,
the National Natural Science Foundation of China (Grant No.~61875182,
No.~11575174, No.~11874346, and No.~11574297),
Anhui Initiative in Quantum Information Technologies, 
and Fundamental Research Funds for the Central Universities (WK2340000083).

\bibliography{RS-MDI-nEB}

\clearpage

\appendix

\begin{widetext}

\section{The RS-MDI certification of non-EB channels}

In this section, we introduce the details of realistic source based measurement-device-independent (RS-MDI) approach proposed and verified in the main text, including the experimental bound for all EB channels, the decoy-state method to exclude multiphoton contributions, and the exact form of payoff functions used in the experiment. We begin by introducing the result of semi-quantum signaling game (SQSG), which was meticulously proposed in the Rosset-Buscemi-Liang's paper \cite{rosset2018resource} (see Fig.1 of the main text). When the referee can perfectly prepare an information complete set of quantum states, a positive value of the average payoff 
\begin{equation}
\label{eq:SM-payoff1}
I_\mathcal{N}=\sum_{x,y,b}\wp(b,x,y)P_\mathcal{N}\left(b|\xi_x,\psi_y\right),
\end{equation}
suggests the ability of the channel $\mathcal{N}$ to convey and maintain entanglement. Here, 
$P_\mathcal{N}\left(b|\xi_x,\psi_y\right)$ is the probability of Abby obtaining $b$ given the referee asks quantum questions $\xi_x$ and $\psi_y$, and $\wp(b,x,y)$ is the payoff function assigned by the referee.
Theoretically, if $\wp(b,x,y)$, $\xi_x$, and $\psi_y$ are properly selected, then for all EB channels, the average payoff is no more than $0$. Thus, a positive value of Eq.~(\ref{eq:SM-payoff1}) certifies that the channel $\cal N$ under test is non-EB. In this work, we further let $\wp(b\neq 0,x,y)=0$ and only consider 
\begin{equation}
\label{eq:SM-payoff}
I_\mathcal{N}=\sum_{x,y}\wp(0,x,y)P_\mathcal{N}\left(b|\xi_x,\psi_y\right).
\end{equation}

\subsection{The experimental bound}\label{practical-bound}

The ideal SQSG assumes that quantum states are perfectly prepared and form an information complete set. In practice, state preparation unavoidably involves flaws and errors due to realistic devices. Also, information complete sets may not be accessible due to limitations on physical systems and devices. In fact, few quantum states may be sufficient for the certification of some quantum channels. Such problems can be solved by applying an experimental bound.

To prove Eq. (2) in the main text, notice that the EB channel ${\cal N}_{\textrm{EB}}$ is in general a measure-and-prepare
channel, i.e., 
\[
{\cal N}_{\textrm{EB}}\left(\rho\right)=\sum_{k}{\rm tr}\left[E_{k}\rho\right]\gamma_{k},
\]
where $\left\{ E_{k}|E_{k}\geqslant0,\sum_{k}E_{k}=\mathbb{I}\right\} $
is a set of POVMs and $\left\{ \gamma_{k}\right\} $ is a set of quantum
states. In the RS-MDI certification, we relax sets of $\left\{\xi_x\right\}$ and $\left\{\psi_y\right\}$ to be sets of arbitrary states. The maximal average payoff for all EB channels ${\cal N}_{\textrm{EB}}$, given the payoff function $\wp\left(0,x,y\right)$, can then be bounded by
\begin{equation}
\begin{aligned}
C_{\textrm{EB}} & =\max_{{\cal N}_{\textrm{EB}}}I_{{\cal N}_{\textrm{EB}}}=\max_{{\cal N}_{\textrm{EB}}}\sum_{x,y}\wp\left(0,x,y\right)P_{{\cal N}_{\textrm{EB}}}\left(0|\xi_{x},\psi_{y}\right)\\
 & =\max_{{\cal N}_{\textrm{EB}},M}\sum_{x,y}\wp\left(0,x,y\right){\rm tr}\left[{\cal N}_{\textrm{EB}}\left(\xi_{x}\right)\otimes\psi_{y}M\right]\\
 & =\max_{E_{k},F_{k}}\sum_{x,y,k}\wp\left(0,x,y\right){\rm tr}\left[E_{k}\xi_{x}\right]{\rm tr}\left[\gamma_{k}\otimes\psi_{y}M\right]\\
 & =\max_{E_{k},F_{k}}{\rm tr}\left[W\sum_{k}E_{k}^{T}\otimes F\left(k\right)^{T}\right],
\end{aligned}
\end{equation}
where $M$ and $F\left(k\right)={\rm tr}_{1}\left[\gamma_{k}\otimes\mathbb{I}M\right]$
are POVMs acting on ${\cal N}_{\textrm{EB}}\left(\xi_{x}\right)\otimes\psi_{y}$
and $\psi_{y}$, respectively. Here, we denote
\begin{equation}
W=\sum_{xy}\wp\left(0,x,y\right)\xi_x^T\otimes\psi_y^T.
\end{equation}

When $\xi_x$ and $\psi_y$ are from information complete sets, and $\wp\left(0,x,y\right)$ are chosen properly, $-W$ can be constructed as an EW for the Choi state of a quantum channel \cite{rosset2018resource,Wilde2013}. In this case, the bound $C_{\textrm{EB}}$ is exactly
$0$ as $E_{k}^{T}\otimes F\left(k\right)^{T}$ are separable positive
operators and ${\rm tr}\left[WE_{k}^{T}\otimes F\left(k\right)^{T}\right]\leqslant0$
holds for all $k$. For arbitrary sets of quantum states $\xi_x $ and $\psi_y$, this bound can also be analytically derived as 
\begin{equation}
C_{\textrm{EB}}^{\textrm{exp}} =\max_{E_{k},F_{k}}{\rm tr}\left[\tilde{W}\sum_{k}E_{k}^{T}\otimes F\left(k\right)^{T}\right] =d^{2}\max_{\omega_{\textrm{sep}}}{\rm tr}\left[\tilde{W}\omega_{\textrm{sep}}\right],
\label{eq:expbound}
\end{equation}
where $\omega_{\textrm{sep}}$ is a separable state.
To see this, notice that $\sum_k E_k ={\mathbb I}$ and $0\leqslant F\left(k\right)\leqslant{\mathbb I}$, $\sum_{k}E_{k}^{T}\otimes F\left(k\right)^{T}$ satisfies
\begin{equation}
\sum_{k}E_{k}^{T}\otimes F\left(k\right)^{T} =d^{2}\sum_{k}\frac{{\rm tr}\left[E_{k}^{T}\right]}{d}\frac{E_{k}^{T}}{{\rm tr}\left[E_{k}^{T}\right]}\otimes\frac{F\left(k\right)^{T}}{d},
\end{equation}
where $E_k^T/{\rm tr}\left[E_k^T\right]$ and $F\left(k\right)^T/d$ can be viewed as quantum states and unnormalized quantum states, respectively.

\subsection{Evaluation of single-photon detections}

Considering the multiphotons in real photon sources, we apply the decoy-state technique to weak coherent pulses (WCPs), such that the single-photon detection events can be efficiently evaluated \cite{hwang2003quantum,wang2005beating,lo2005decoy}.
Generally, the quantum states of phase-randomized WCPs can be written in the Fock basis as
\begin{equation}
\rho_{\alpha}=e^{-\alpha}\sum_{n=0}^\infty\frac{\alpha^{n}}{n!}\left|n\right\rangle \left\langle n\right|,
\end{equation}
where $\alpha$ is the mean photon number per pulse and $n$ is the photon number. When pulses $\xi_{x}$ and $\psi_{y}$ are prepared with intensities $\alpha_{\xi}$ and $\alpha_{\psi}$, respectively, the probability to obtain the result $b$ of a jointly measurement can be written as 
\begin{equation}
\label{eq:gain}
Q_{b,\xi_{x},\psi_{y}}^{\alpha_{\xi}\alpha_{\psi}}=e^{-\alpha_{\xi}-\alpha_{\psi}}\sum_{n,m=0}^{\infty}\frac{\alpha^{n}_{\xi} \alpha^{m}_{\psi}}{n! m!}Y^{nm}_{b,\xi_{x},\psi_{y}}.
\end{equation}
Here, $Q_{b,\xi_{x},\psi_{y}}^{\alpha_{\xi}\alpha_{\psi}}$ is the ratio of the number of detection events to the number of emitted pulse pairs in $\xi_{x}$ and $\psi_{y}$ with mean photon numbers $\alpha_{\xi}$ and $\alpha_{\psi}$, respectively. Consequently, $Y^{nm}_{b,\xi_{x},\psi_{y}}$ is the conditional probability of detection events $b$ given that $n$-photon and $m$-photon pulses are emitted in $\xi_{x}$ and $\psi_{y}$, respectively \cite{mdiqkd,Xu2013practical}. 

The decoy-state method is applied when $\xi_x$ and $\psi_y$ are randomly prepared with different intensities. 
We use three types of mean photon per pulse values, i.e., $\alpha_{\xi},\alpha_{\psi}\in\left\{ \mu,\nu,\omega | \mu>\nu>\omega=0 \right\}$, and obtain seven gains of $Q_{b,\xi_{x},\psi_{y}}^{\alpha_{\xi_x}\alpha_{\psi_y}}$ with $\{\alpha_{\xi}\alpha_{\psi}\}$ $\in$ $\{\mu\mu, \nu\nu, \mu\omega, \omega\mu, \nu\omega, \omega\nu, \omega\omega\}$. Let $J_{1}$ and $J_{2}$ be ($\xi_{x}$, $\psi_{y}$, and $b$ are omitted for simplicity), 
\begin{equation}
\begin{aligned}
J_{1} & = Q_{\nu\nu}e^{2\nu}+Q_{\omega\omega}-Q_{\nu\omega}e^{\nu}-Q_{\omega\nu}e^{\nu},\\
J_{2} & = Q_{\mu\mu}e^{2\mu}+Q_{\omega\omega}-Q_{\mu\omega}e^{\mu}-Q_{\omega\mu}e^{\mu}.
\end{aligned}
\end{equation}
We calculate the equation $\mu^{3}J_{1}-\nu^{3}J_{2}$, where $Y^{n,0},Y^{0,m}$ and $Y^{1,2},Y^{2,1}$ can be canceled, and obtain
\begin{equation}
Y^{11}=  \frac{\mu^{3}J_{1}-\nu^{3}J_{2}}{\mu^{2}\nu^{2}\left(\mu-\nu\right)} -\frac{1}{\mu^{2}\nu^{2}\left(\mu-\nu\right)}\sum_{n,m \geqslant 2}\frac{\mu^{3}\nu^{n+m}-\nu^{3}\mu^{n+m}}{n!m!}Y^{nm}.
\end{equation}
Since $Y^{n,m}\in[0,1]$ and $\mu>\nu$, the lower bound $Y^{1,1,L}$ and upper bound $Y^{1,1,U}$ can be written as
\begin{equation}
\begin{aligned}
Y^{11,L}= & \frac{\mu^{3}J_{1}-\nu^{3}J_{2}}{\mu^{2}\nu^{2}\left(\mu-\nu\right)},\\
Y^{11,U}= & Y^{11,L} -\frac{\mu^3 \left( e^\nu-1-\nu \right)^2 - \nu^3 \left( e^\mu-1-\mu \right)^2}{\mu^{2}\nu^{2}\left(\mu-\nu\right)},
\end{aligned}
\end{equation}
respectively.

\subsection{The 6-state and 4-state average payoff}
In this work, we consider two kinds of average payoff with different sets of input states.

\subsubsection{The 6-state average payoff}
For the first kind, we use an information complete set of states, i.e., eigenstates of three Pauli matrices denoted as $\xi_{x},\psi_{y}$ $\in\left\{ \left|0\right\rangle _{Z},\left|1\right\rangle _{Z},\allowbreak \left|0\right\rangle _{X},\left|1\right\rangle _{X},\left|0\right\rangle _{Y},\left|1\right\rangle _{Y} \right\}$, respectively.
The corresponding payoff function is 
\begin{equation}
\label{six-state-payoff}
\wp\left(0,\xi_{x},\psi_{y}\right)=
\begin{cases}
\frac{1}{4}, & \left(\xi_{x},\psi_{y}\right) \text{ anti-correlated in } X \text{ or } Z;\\
-\frac{1}{4}, & \left(\xi_{x},\psi_{y}\right) \text{ correlated in } X \text{ or } Z;\\
-\frac{1}{2}, & \left(\xi_{x},\psi_{y}\right)\text{ correlated in } Y ;\\
0, & \text{otherwise}.
\end{cases}
\end{equation}
Here, ``anti-correlated'' represents $\left(0,1\right)$ or $\left(1,0\right)$ in the respective bases and ``correlated'' represents $\left(0,0\right)$ or $\left(1,1\right)$ in the respective bases.

By replacing $P_{\cal N}(b|\xi_x,\psi_y)$ with $Y^{1,1,L}$ or $Y^{1,1,U}$ according to the sign of $\wp(0,\xi_x,\psi_y)$,  the lower bound of the payoff value $I_{\cal N}$ can be written as
\begin{equation}
\begin{aligned}
I_{\cal N} \geqslant I_{\cal N}^L= & -\frac{1}{4} Y^{11,U}_{0_Z 0_Z} + \frac{1}{4} Y^{11,L}_{0_Z 1_Z}
+ \frac{1}{4} Y^{11,L}_{1_Z 0_Z} -\frac{1}{4} Y^{11,U}_{1_Z 1_Z}\\
& -\frac{1}{4} Y^{11,U}_{0_X 0_X} + \frac{1}{4} Y^{11,L}_{0_X 1_X}
+ \frac{1}{4} Y^{11,L}_{1_X 0_X} -\frac{1}{4} Y^{11,U}_{1_X 1_X}\\
&-\frac{1}{2} Y^{11,U}_{0_Y 0_Y} -\frac{1}{2} Y^{11,U}_{1_Y 1_Y}.
\end{aligned}
\end{equation}

When six states are perfectly prepared, the average payoff from Eq. (\ref{six-state-payoff}) has the maximal value $0.5$, achieved by the identity channel. 
The corresponding ideal EB bound is $0$, which can also be proven by Eq. (\ref{eq:expbound}).
In our experiment, based on the assumption that the referee has full knowledge of his states, he can use state tomography to determine the exact density matrices of the prepared states (as shown in Sec. \ref{state-tomo}). 
Then, the experimental EB bound is numerically calculated as $0.047$, higher than the ideal bound of $0$.
Also, the maximal value of $I_{\cal N}$ is calculated as 0.451, which is slightly lower than $0.5$ of the ideal case due to inaccurate state preparations.

\subsubsection{The 4-state average payoff}  

For the second kind, we use an information \textit{incomplete} set of states, i.e., eigenstates of the Pauli matrices $Z$ and $Y$, denoted as $\xi_{x},\psi_{y}$ $\in\left\{ \left|0\right\rangle _{Z},\left|1\right\rangle _{Z},\allowbreak \left|0\right\rangle _{Y},\left|1\right\rangle _{Y} \right\}$, respectively.
The corresponding payoff function is 
\begin{equation}
\label{four-state-payoff}
\wp\left(0,\xi_{x},\psi_{y}\right)=
\begin{cases}
\frac{1}{4}, & \left(\xi_{x},\psi_{y}\right) \text{ anti-correlated in }  Z;\\
-\frac{1}{4}, & \left(\xi_{x},\psi_{y}\right) \text{ correlated in }  Z;\\
-\frac{1}{2}, & \left(\xi_{x},\psi_{y}\right)\text{ correlated in } Y ;\\
0, & \text{otherwise}.
\end{cases}
\end{equation}

According to the sign of $\wp(0,\xi_x,\psi_y)$ in Eq. (\ref{four-state-payoff}),  the lower bound of the payoff value $I_{\cal N}$ can be written as
\begin{equation}
\begin{aligned}
I_{\cal N} \geqslant I_{\cal N}^L= & -\frac{1}{4} Y^{11,U}_{0_Z 0_Z} + \frac{1}{4} Y^{11,L}_{0_Z 1_Z}
+ \frac{1}{4} Y^{11,L}_{1_Z 0_Z} -\frac{1}{4} Y^{11,U}_{1_Z 1_Z}\\
&-\frac{1}{2} Y^{11,U}_{0_Y 0_Y} -\frac{1}{2} Y^{11,U}_{1_Y 1_Y}.
\end{aligned}
\end{equation}

When four states are perfectly prepared, the average payoff from Eq. (\ref{four-state-payoff}) has the maximal value $0.25$, and the ideal EB bound is $0$.
In our experiment, the experimental EB bound can be calculated as $0.025$, and the maximal value is $0.226$.

\section{Experiment details}
 
In this section, we introduce the basic techniques of the experiments, including feedback systems, the design and implementation of the decoherence channel, secure key rates using the noise-added fiber channel, and the quantum state tomography.

\subsection{Setup details and feedback systems}

Our experimental setup is composed of three portions: state preparation, detection, and the channel to be tested. 
For state preparation, DFB1 (DFB2) at wavelength of $1550.12$ nm is used and directly modulated to emit pulses of $37.5$ MHz repetition rate, which are narrowed to $2.5$ ns at FWHM using an IM. 
Time-bin states are created using an AMZI that separates the pulses at $6.5$ ns, and the basis of $Z$ or $X$($Y$) is chosen with the following IM. 
Phase states are created using the FR, PM, and CIR, where the optical pulses travel through the PM twice for lower modulation voltage. 
In addition, another IM is used to adjust the intensity difference between pulses of $Z$ and $X$($Y$) bases. 
All IMs and PMs for preparing $\xi_x$ and $\psi_y$ are controlled by independent random numbers. 
The pulses are lowered down to single-photon level with an EVOA, and are filtered with a $100$ GHz narrow pass-band filter for spectral noise. 

In addition, we adopt feedback systems for phase, optical intensity, wavelength, and bias voltage
to ensure the stability of the entire system.

In order to obtain high interference visibility in the $X$ and $Y$ basis at the Bell-state measurement (BSM), a phase reference needs to be established between the two state preparation modules, i.e., the phase difference between the two asymmetric Mach-Zehnder interferometers (AMZI) should be $2\pi k$, with $k$ a positive integer. Pulses of wavelength $1550.12$ nm and frequency of $25$ MHz are sent from DFB3 through the two AMZI in sequence, as shown in Fig. 2 in the main text. Here, the pulses can be detected in three time-windows, where only those in the second time window show interference. A gated InGaAs single-photon detector (SPD) is used to detected the interference photons, and the $2\pi$ phase drift of the AMZIs is measured to be $\sim8$ minutes. For the experiment, a real-time phase feedback system is built. A fiber phase shifter is placed on the short arm of AMZI1 to maintain the interference steady at destructive interference, i.e., the interference photons are monitored with the SPD to remain minimum photon count. Thus, the phase reference between $\xi_{x}$ and $\psi_{y}$ is established.

For optimal BSM, firstly the arrival time of both $\xi_{x}$ and $\psi_{y}$ pulses are calibrated at the BSM site. To compensate the difference in arrival time to the BSM for pulses $\mathcal{N}(\xi_{x})$ and $\psi_{y}$, DFB1 and DFB2 alternatively send pulses to the SNSPDs. 
Based on arrival time difference between the two pulses, and using a programmable delay chip to adjust the pulse delays, $\mathcal{N}(\xi_{x})$ and $\psi_{y}$ pulses are precisely overlapped at the BS. 
The detection efficiency is $27\%$ and the dark count rate is $50$ counts per second for each SNSPD. 
To achieve optimal trade-off between effective coincidence count rates and eliminating imperfect phase encodings at the pulse edges of $X$ and $Y$ basis, the effective BSM time window is set to $~85\%$. Secondly, the wavelengths of both $\xi_{x}$ and $\psi_{y}$ are required to be almost the same. Here, the temperature of DFB2 is scanned, and the Hong-Ou-Mandel dip is measured. The temperature is set to the value where minimum coincidence count of $55\%$ occurs. Thus, the wavelengths are optimal. Thirdly, considering the fluctuations of the source and bias voltages of IMs, the intensities of $\xi_{x}$ and $\psi_{y}$ pulses are calibrated every $\sim600$ seconds. 

The clock reference for the entire system is established by sending synchronization laser pulses at $1570$ nm of $500$ kHz through a separate fiber to each of $\xi_x$ and $\psi_y$ state preparation modules. 
Then, they are detected through a photodiode and the system repetition rate of $37.5$ MHz is generated. 
With this configuration, we optimized pulse modulation from the modulators such that fast real-time data acquisition is realized.

\subsection{Realization of the decoherence channel through the Sagnac interferometer}

As discussed in the main text, the decoherence channel
preserves populations of the state in the $Z$ basis, while the coherence between them, i.e., the relative phases, is suppressed. 
To realize such a channel, we design the Sagnac interferometer (SI) with a phase modulator (PM) to randomly eliminate the first or second time-bin when states in $X$ and $Y$ bases are prepared.
Initially, four input states $\left| 0 \right\rangle_X$, $\left| 1 \right\rangle_X$, $\left| 0 \right\rangle_Y$, $\left| 1 \right\rangle_Y$ are encoded in the phase between first and second time-bins, written as $|\sqrt{\alpha/2}\rangle ^{s} |e^{i\phi}\sqrt{\alpha/2}\rangle ^{l}$ with $\phi = 0,\pi,\pi/2$, $3\pi/2$ for simplicity. Here, $s$ and $l$ represent the short and long arm of the AMZI, respectively.

As shown in Fig.~2b of the main text, when entering the SI, $\xi_x$ is firstly split into two paths $a$ and $b$ by the beam splitter (BS), with path $a$ a shorter arrival time to the PM than that of path $b$. Then, $\xi_x$ is split into four pulses. 
By precisely adjusting the length of paths $a$ and $b$, we set the arrival time for the pulses of two paths in a difference of $13.3$ ns. Now, the time-bins in the sequence of $|\sqrt{\alpha}/2\rangle _{a}^{s},|e^{i\phi}\sqrt{\alpha}/2\rangle _{a}^{l},|\sqrt{\alpha}/2\rangle _{b}^{s},|e^{i\phi}\sqrt{\alpha}/2\rangle _{b}^{l}$ are modulated by the PM, adding relative phases $\theta_{a}^{s},\theta_{a}^{l},\theta_{b}^{s},\theta_{b}^{l}$. Below we write the exact process for $\xi_x$ to transmit through the SI,
\begin{small}
\begin{equation*}
\begin{aligned}
\left|\sqrt{\frac{\alpha}{2}}\right\rangle ^{s}\left|e^{i\phi}\sqrt{\frac{\alpha}{2}}\right\rangle ^{l}
&\overset{BS}{\rightarrow}\left|i\frac{\sqrt{\alpha}}{2}\right\rangle _{a}^{s}\left|ie^{i\phi}\frac{\sqrt{\alpha}}{2}\right\rangle _{a}^{l}\left|\frac{\sqrt{\alpha}}{2}\right\rangle _{b}^{s}\left|e^{i\phi}\frac{\sqrt{\alpha}}{2}\right\rangle _{b}^{l}
\overset{PM}{\rightarrow}\left|ie^{i\theta_{a}^{r}}\frac{\sqrt{\alpha}}{2}\right\rangle _{a}^{s}\left|ie^{i\left(\phi+\theta_{a}^{s}\right)}\frac{\sqrt{\alpha}}{2}\right\rangle _{a}^{l}\left|e^{i\theta_{b}^{r}}\frac{\sqrt{\alpha}}{2}\right\rangle _{b}^{s}\left|e^{i\left(\phi+\theta_{b}^{s}\right)}\frac{\sqrt{\alpha}}{2}\right\rangle _{b}^{l}\\
&\overset{BS}{\rightarrow}\left| \left(e^{i\theta_{b}^{r}}-e^{i\theta_{a}^{r}} \right)\frac{\sqrt{\alpha}}{2\sqrt{2}}\right\rangle _{c}^{s}
\left| e^{i\phi} \left(e^{i\theta_{b}^{s}}-e^{i\theta_{a}^{s}} \right) \frac{\sqrt{\alpha}}{2\sqrt{2}}\right\rangle_{c}^{l}
\left|i  \left(e^{i\theta_{a}^{r}}+e^{i\theta_{b}^{r}} \right) \frac{\sqrt{\alpha}}{2\sqrt{2}}\right\rangle_{d}^{s}
\left|i e^{i\phi} \left(e^{i\theta_{a}^{s}}+e^{i\theta_{b}^{s}} \right)\frac{\sqrt{\alpha}}{2\sqrt{2}}\right\rangle_{d}^{l}.
\end{aligned}
\end{equation*}
\end{small}

In the experiment, port $d$ is used for the input to the BSM. By adjusting the modulation voltages on the PM for $\theta_{a}^{s},\theta_{a}^{l},\theta_{b}^{s},\theta_{b}^{l}$, the channel can either model the identity channel or the fully decoherence channel.
Precisely, the identity channel is realized when the PM is turned off, i.e., adding phases $0,0,0,0$ in the same sequence. In this case, the state remains the same and is output in port $d$.
As for the fully decoherence channel, we randomly add phases in the sequence $0,0,0,\pi$ or $0,0,\pi,0$ with the same probability, such that the second or first time-bin is eliminated, respectively. The output of port $d$ is thus in the form of $|\sqrt{\alpha/2}\rangle^s$ or 
$|\sqrt{\alpha/2}\rangle^l$ 
(up to an overall phase), respectively.
By using an independent random number string to control the PM, the identity channel and fully decoherence channel can be realized with probabilities $1-\gamma $ and $\gamma$, respectively. In this manner, decoherence channel of the form 
\begin{equation}
\mathcal{D}_{\gamma}(\rho)=(1-\gamma)\rho+\gamma(|0\rangle\langle0|_{Z} \rho_{00}+|1\rangle \langle1|_{Z}\rho_{11}),
\end{equation}
can be constructed.

\subsection{Secure key rates using the noise-added fiber channel}

Since effects from noise on actual channels are extremely complicated and related to specific channel structures, application environment, detection process etc., for simplicity and without loss of generality, we design the additive noise in a fiber channel by combining photons of a continuous-wave (CW) source with the WCPs of $\xi_x$ into the untrusted measurement, as shown in Fig.~2c of the main text. 
We adjust the intensity of the CW, such that different strengths of noise can be modeled.
We use the ratio of CW intensity to $\xi_x$ pulse intensity per second, denoted by $\beta$, to describe the strength of noise.

Since directly analyzing the effect of noise photons is extremely complicated, here, we use the secure key rate of quantum key distribution to confirm the correct certification of the tested channel against noise.
We suppose that $\xi_x$ and $\psi_y$ are prepared by two distinct users, namely Alice and Bob, and the BSM is performed by an untrusted third party Charlie in the usual measurement-device-independent quantum key distribution scenario. For both Alice and Bob, pulses of $Z$ basis with intensity $\mu$ are used for key generation, and pulses of $X$ basis are used for error estimation. The key rate for measurement-device-independent quantum key distribution is calculated by \cite{mdiqkd},
\begin{equation}\label{eq:keyrate}
R \geq Q^{11}_{Z,\mu\mu}(1 - H(e^{11}_X)) - Q^{\mu\mu}_Z f H(E^{\mu\mu}_Z).
\end{equation}
Here, $f$ is the error correction inefficiency factor, and $H(x) = -x \textrm{log}_2 x -(1-x) \textrm{log}_2(1-x)$ is the binary entropy function. $Q^{11}_{Z,\mu\mu}$, $Q^{\mu\mu}_Z$, $E^{\mu\mu}_Z$, and $e^{11}_X$ are the single photon gain, the total gain, the quantum bit error rate in the $Z$ basis, and the quantum bit error rate in the $X$ basis when $\xi_x$ and $\psi_y$ are both of single photons, respectively \cite{mdiqkd}. The experimental key rates in Fig.~4 of the main text are obtained by taking the measured $Q^{\mu\mu}_Z$ and $E^{\mu\mu}_Z$ into Eq.~(\ref{eq:keyrate}). The theoretical key rate (dotted) line is obtained by simulating Eq.~(\ref{eq:keyrate}) with our experimental parameters, where noise photons from the CW are treated as dark counts \cite{Mao2018QKD}. 

\subsection{Robustness against flawed state preparation}\label{state-tomo}

As discussed in Sec. \ref{practical-bound}, to avoid falsely witnessing non-EB channels, the experimental bound is determined by the full knowledge of quantum questions. The referee can achieve this by performing state tomography of $\xi_{x}$ and $\psi_{y}$.
Generally, the density matrix of an arbitrary qubit state is in the form of
\begin{equation}
\rho = \frac{\mathbb{I}+\left\langle X\right\rangle X +\left\langle Y\right\rangle Y +\left\langle Z\right\rangle Z }{2},
\end{equation}
where $X$, $Y$, $Z$ are the three Pauli matrices, and $\left\langle X\right\rangle$, $\left\langle Y\right\rangle$, $\left\langle Z\right\rangle$ are expectation values when measuring the corresponding Pauli matrices. 
In this experiment, the value $\left\langle Z\right\rangle$ is evaluated by a time-bin intensity ratio $r$, i.e., the ratio of the photon counts in the first time-bin window to that in the second.
To obtain $\left\langle X\right\rangle$ for each $\xi_{x}$ and $\psi_{y}$, the corresponding pulses prepared by one AMZI are sent through the other AMZI. 
Photon pulses can be detected at both output ports of the final BS, where they are observed in three consecutive time windows using a gated SPD.
Interference is shown in the second time window, where the phase differs $\pi$ for two ports.
By using the photon counts of three time window in both BS outputs, values of $\left\langle X\right\rangle$ can be obtained.
For the value of $\left\langle Y\right\rangle$, we adjust the phase shifter of AMZI1 (see Fig.~2a of the main text), such that two output ports of the final BS correspond to the projection onto relative phases $\pi/2$ and $3\pi/2$ \cite{Sun2016}.

The constructed density matrices of $\tilde{\xi}_x$ and $\tilde{\psi}_y$ are taking into Eq.~(\ref{eq:expbound}), such that the experimental bound for all EB channels are determined. Here, we show the tomography results in Fig. \ref{fig:tomo} and list fidelity \cite{nielsen2002quantum} of our input states in Table \ref{tab:fidelity}.

\begin{figure}\centering
\includegraphics[width=\linewidth]{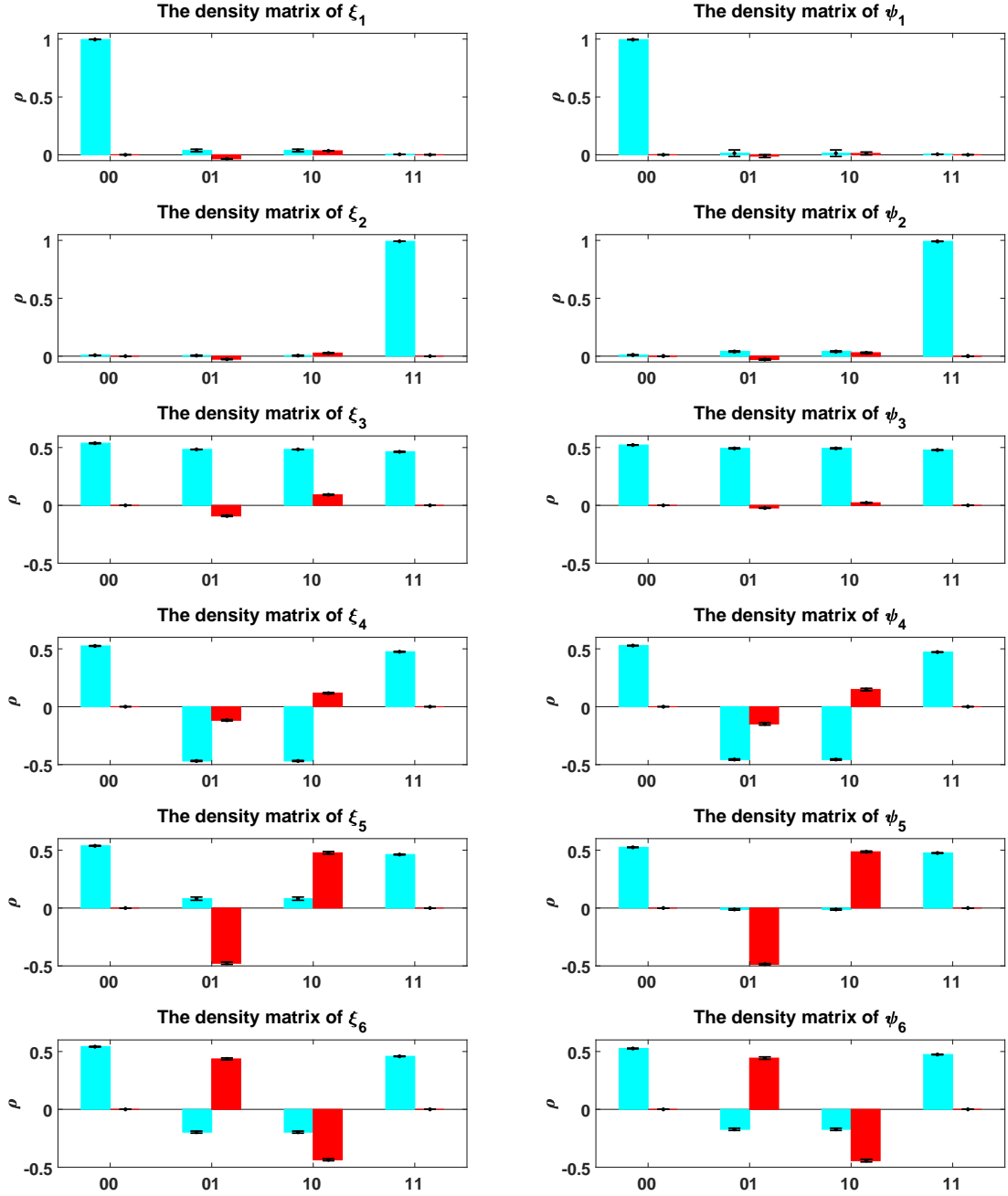}
\caption{\label{fig:tomo}The tomography results of quantum states used in this experiment. 
In each figure, four blue bars represent the real part of four entries in the density matrix, while the red bars represent the imaginary part. Based on this description of the quantum states, the referee can determine the bound $C_{EB}$ of all EB channels.}
\end{figure}

\begin{table}[h]
\centering
\caption{Fidelity of input quantum states}
\begin{tabular}{c|cccccc}
\hline 
\text{Fidelity}(\%) & $\left|0\right\rangle_Z$  & $\left|1\right\rangle_Z$  & $\left|0\right\rangle_X$  & $\left|1\right\rangle_X$  & $\left|0\right\rangle_Y$  & $\left|1\right\rangle_Y$ \\
\hline $\xi_{x}$ & 99.7 & 99.2 & 98.4 & 96.9 & 97.7 & 93.6\\
$\psi_{y}$ & 99.5 & 99.1 & 99.3 & 95.7 & 98.6 & 94.3\\
\hline 
\end{tabular}
\label{tab:fidelity}
\end{table}

%
%
\end{widetext}

\end{document}